\documentclass[aps,pra,11pt,superscriptaddress]{revtex4-2}
\usepackage{amsthm}  % Load amsthm first
\usepackage{amsmath,amssymb,amsfonts,graphicx,bbm,mathtools,braket,color,dsfont}

\usepackage{array, graphicx, dcolumn,mathrsfs}
\usepackage{bm,diagbox,natbib, xcolor} 

\newtheorem{theorem}{Theorem}[section]
\newtheorem{corollary}[theorem]{Corollary}

\newcommand{\ip}[2]{\left\langle{#1}\right|\left.\!{#2}\right\rangle}
\DeclareMathOperator{\Tr}{Tr}
\newcommand{\abs}[1]{\left|#1\right|}  
\newcommand{\detQ}{\det [Q]}  
\newcommand{\mga}[1]{{#1}}
\newcommand{\jh}[1]{{#1}}
\newcommand{\gf}[1]{{#1}}
\begin{document}
\title{Orders matter: tight bounds on the precision of sequential quantum estimation for multiparameter models}
%\title{\mga{Three possible titles here: 1. Beyond Holevo: stepwise estimation for multiparameter quantum metrology; 2. Quantum multiparameter estimation: when stepwise strategies outperform joint measurements; 3. Orders matter: tight bounds on the precision of sequential quantum estimation for multiparameter models}}

\author{Gabriele Fazio}
\email{gabriele.fazio@studenti.unimi.it (G. Fazio)}
\affiliation{Dipartimento di Fisica Aldo Pontremoli, Università degli Studi di Milano, I-20133 Milano, Italy}
\author{Jiayu He} 
\email{jiayu.he@helsinki.fi (J. He)}
\affiliation{QTF Centre of Excellence, Department of Physics, University of Helsinki, FI-00014 Helsinki, Finland}
\author{Matteo G. A. Paris}
\email{matteo.paris@fisica.unimi.it (M. G. A. Paris)}
\affiliation{Dipartimento di Fisica Aldo Pontremoli, Università degli Studi di Milano, I-20133 Milano, Italy}
\date{\today}
\begin{abstract}
In multiparameter quantum metrology, the ultimate precision of joint estimation is dictated by the Holevo 
Cramér-Rao bound. \mga{In this paper, we discuss and analyze in detail an alternative approach:} the stepwise estimation strategy. In this approach, parameters are estimated sequentially, using an optimized fraction of the total available resources allocated to each step. We derive a tight and achievable precision bound for this protocol, the stepwise separable bound, and provide its closed-form analytical expression, revealing a crucial dependence on the chosen measurement ordering. We provide a rigorous comparison with the joint measurement strategy, deriving analytical conditions that determine when the stepwise approach offers superior precision. Through the analysis of several paradigmatic SU(2) unitary encoding models, we demonstrate that the stepwise strategy can indeed outperform joint measurements, particularly in scenarios characterized by \mga{non-optimal probes or models with a high degree of sloppiness. Our findings establish stepwise estimation as a powerful alternative to joint and collective measurements, proving that sequential protocols can provide a genuine metrological advantage, especially in resource-constrained or imperfect experimental settings.}
\end{abstract}
\maketitle

\section{Introduction}
Quantum metrology seeks to leverage uniquely quantum features such as entanglement, coherence, and squeezing to surpass the precision limits imposed by classical strategies \cite{giovannetti2004,paris2009quantum,montenegro24}. 
While significant theoretical and experimental progress has been made in single-parameter quantum estimation, the simultaneous estimation of multiple parameters has emerged as a challenging frontier. Multiparameter quantum estimation underpins a broad spectrum of applications \cite{szczykulska2016multi,wang20,albarelli2020perspective,di2022multiparameter,montenegro24}, from high-resolution imaging \cite{chrostowski2017super,lee2022quantum} to tests of fundamental physics \cite{gupta2020multiparameter,ghosh2019weak}. 
However, it also introduces conceptual and technical challenges: the optimal measurements for different 
parameters may not commute, rendering it impossible to simultaneously saturate the individual quantum 
Cramér–Rao bounds (QCRBs)\cite{heinosaari2016,ragy2016,dimensionMatter}.
In such cases, the ultimate limit on precision is dictated by the Holevo bound, which remains tight but notoriously difficult to evaluate and saturate\cite{holevo2011}. \jh{It characterizes the ultimate achievable precision in the asymptotic regime, \mga{but it typically requires collective measurements} over infinitely many copies of the quantum probe. Nagaoka introduced an alternative bound based on separable measurements \cite{Nagaoka2005}. Although generally weaker than the Holevo bound, it may be achieved using sepaarate (non collective) measurement schemes, and it has been shown to coincide with the Holevo bound in specific cases, such as two-parameter estimation with pure states in a two-dimensional Hilbert space \cite{suzuki}. }

\mga{However, both the Holevo and Nagaoka bounds lack closed-form analytic expressions in the general case, which limits their utility as benchmarks and in comparing strategies. To bridge the gap between theoretical bounds and practical feasibility, a hierarchy of precision limits has been developed. Among these, the $R$ \cite{carollo2019,razavian2020quantumness} and $T$ \cite{HeWeight} bounds stand out for their analytic tractability and close connection to the geometry of quantum states, i.e., the quantum Fisher information matrix  and the mean Uhlmann curvature. These bounds provide a useful bracketing of the Holevo bound, clarifying how quantum incompatibility and the sloppiness of model \cite{PRXQuantum.2.020308,machta2013parameter,Yang23,PhysRevLett.97.150601,sloppiness,Bizzarri25,frigerio2024overcoming}, i.e., the possible inefficiency of the parameter encoding, limit the achievable precision.}

\jh{Motivated by these considerations, we discuss in details a novel approach, and its corresponding precision bound, that is easier to obtain in analytic form and has operational relevance.} In this context, we consider a stepwise estimation protocol \cite{mukhopadhyay2025beating}, in which different parameters are estimated sequentially by allocating resources to different subsets of the measurement data. Such a strategy offers a compromise between the simplicity of fully separable single-parameter estimation and globally joint multiparameter estimation, while providing a tractable analytical framework. This is particularly relevant in scenarios where joint strategies are impractical \cite{yang2025overcoming} or when experimental constraints limit the accessibility \gf{to} optimal probes \cite{sharma2025mitigating}. 

In this work, we develop a general framework for stepwise estimation in quantum multiparameter metrology. We introduce a class of precision bounds tailored to such sequential strategies and investigate their dependence on parameter ordering, as well as their relation to existing bounds such as the Holevo, symmetric logarithmic derivative (SLD) QCRB and $T$/$R$-type bounds. Our results demonstrate that stepwise estimation can not only provide operationally meaningful bounds but may, under certain conditions, outperform  joint strategies, particularly in the presence of sloppiness, i.e., when only certain combinations of parameters affect the quantum state significantly \cite{parameterHierarchy}, or under resource constraints that prevent the preparation of ideal probe states.

The paper is organized as follows. In Section II, we review the general framework of quantum multiparameter estimation, including the definitions of the classical and quantum Cram\`er-Rao bounds and the Holevo bound, and discuss their interrelations and attainability conditions. Section III introduces the concept of stepwise estimation strategies and formulates a family of precision bounds associated with sequential resource allocation. We derive closed-form analytic expressions for these bounds, establish their optimality with respect to different parameter orderings, and provide inequalities that characterize their performance relative to known quantum bounds. Section IV illustrates the application of our framework to explicit physical models, including two- and three-parameter estimation scenarios with qubit and qutrit probes \cite{dimensionMatters,pal2025role}. These examples highlight conditions under which stepwise estimation outperforms joint estimation, especially when the quantum Fisher information is highly anisotropic or the parameters are incompatible. Finally, Section V summarizes our main findings and discusses possible future directions.

\section{Framework of Multiparameter quantum estimation}
In this Section, we outline the theoretical background for multiparameter quantum estimation. Let $\rho_{\bm \lambda}$ be a quantum state on a finite-dimensional Hilbert space, parameterized by a vector of $n$ real parameters $\bm\lambda=(\lambda_1,\ldots, \lambda_n)^T$, and  $\left\{\Pi_k\right\}$ with $\Pi_k\geq0$ and $\sum_{k} \Pi_k=\mathbb{I}$ a positive operator-valued measurement (POVM). The probability of obtaining outcome $k$ is given by $p_{\bm \lambda}(k) = \Tr \left[\rho_{\bm \lambda}\Pi_k\right]$. A corresponding estimator $\hat{\bm\lambda}(k)$ is assigned to each outcome, and the performance is evaluated via the covariance matrix $V(\hat{\bm\lambda})$, whose components read:
\begin{equation}
V_{\mu\nu} =\sum_k p_{\bm \lambda}(k)[\lambda_\mu(k)-E_k(\hat \lambda_\mu)][ \lambda_\nu(k)-E_k(\hat \lambda_\nu)].
\end{equation}
where $E_k(\hat \lambda_\mu)$ denotes the expectation value of the estimator $\hat{\lambda}_\mu$ under the distribution $p_{\bm\lambda}(k)$.

Assuming locally unbiased estimators, i.e. $E[\hat{\lambda}_\mu] = \lambda_\mu$ and $\partial_\nu E(\hat{\lambda}_\mu) = \delta_{\mu\nu}$, the classical Cramér-Rao bound (CRB) provides a fundamental lower limit on the achievable covariance\cite{cramer1999}:
\begin{equation}
 V(\hat{\bm \lambda})\geq \frac{1}{M}F^{-1}\,,
\end{equation}
with $F$ the Fisher information matrix (FIM), and $M$ the number of repeated measurements. The elements of FIM are defined as:
\begin{equation}
F_{\mu\nu} \equiv \sum_k \frac{\partial_\mu  p_{\bm \lambda}(k)\, \partial_\nu  p_{\bm \lambda}(k)}{p_{\bm \lambda}(k)}
\,.
\end{equation}
The CRB can be saturated in the asymptotic limit of an infinite number of repeated experiments using Bayesian or maximum likelihood estimators \cite{kay1993}. 

In the quantum setting, due to non-commutativity of observables, multiple versions of quantum Fisher information matrices (QFIMs) arise. One of the most prominent among them is defined through the symmetric logarithmic derivatives (SLDs) $L_\mu^S$ \cite{helstrom1967}, defined as operators which satisfy:
\begin{equation}
\partial_\mu\rho_{\bm \lambda}=\frac{L_\mu^S\rho_{\bm \lambda}+\rho_{\bm \lambda} L_\mu^S}{2}
\end{equation}
The \gf{SLD-}QFIM is then defined as
\begin{equation}
Q_{\mu \nu}\equiv \frac{1}{2}\Tr\left[\rho_{\bm \lambda}\{L_\mu^S,L_\nu^S\}\right]\,,
\end{equation}
where $\{A,B\}=AB+BA$ denotes the anti-commutator of the operators $A$ and $B$.
In the case of pure statistical models,  $\rho_{\bm\lambda} = |\psi_{\bm\lambda}\rangle \langle \psi_{\bm\lambda}|$, the QFIM reduces to:
\begin{align*}
Q_{\mu\nu}  = 4\,\textrm{Re}\big(\ip{\partial_\mu\psi_{\bm\lambda}}{\partial_\nu\psi_{\bm\lambda}}-\ip{\partial_\mu\psi_{\bm\lambda}}{\psi_{\bm\lambda}}\ip{\psi_{\bm\lambda}}{\partial_\nu\psi_{\bm\lambda}}\big),
\end{align*}
where $\partial_k\equiv\partial_{\lambda_k}$.
%%%
\subsection{\jh{ General Quantum Cramér-Rao Bounds}}
\jh{In this subsection, we present a general framework for quantum Cramér-Rao bounds (QCRBs), covering all commonly uesd quantum bounds, including the SLD bound $C_\text{S}$, the Holevo bound $C_\text{H}$, as well as the $R$ and $T$ bounds, and their hierarchical relationships.}

Utilizing the QFIM one can formulate scalar bounds for estimation error under a given cost matrix $W$ (a real, positive definite $d \times d$ matrix). The quantum Cramér-Rao Bound (QCRB), \gf{or SLD QCRB} states that $\text{Tr}[V(W,\hat{\bm \lambda})] \geq C_\text{S}[W,\hat{\bm\lambda}]$, with
\begin{equation}
C_\text{S}[W,{\hat{\bm \lambda}}] \equiv\frac1M\,\Tr\left[ W\, Q^{-1}\right],
\end{equation}
\jh{where $M$ is the number of independent repetitions of measurement. Increasing it can reduce statistical errors.} 
However, due to non-commuting SLDs, the bound isn't always tight. A more fundamental limit is provided by the Holevo bound\cite{holevo2011,albarelli2019evaluating}:
\begin{equation}
C_\text{H}[ W,{\hat{\bm \lambda}}]
\equiv \min_{X \in \mathcal X} \Big\{\Tr\left[ W\, \textrm{Re}\left( Z[X]\right)\right]+\Tr \left[\left|\left|W\, \textrm{Im}\left( Z[ X]\right)\right|\right|_1\right]\Big\},
\end{equation}
where $Z_{\mu\nu} \equiv \Tr[\rho_{\bm\lambda} X_\mu X_\nu]$, and the set $\mathcal{X}$ contains Hermitian operators $X_\mu$ satisfying the local unbiased condition: $\Tr[\partial_\nu \rho_{\bm\lambda} X_\mu] = \delta_{\mu\nu}$. The Holevo bound becomes asymptotically achievable in the limit of collective measurements over many copies of the state\cite{hayashi2008,kahn2009,Yamagata2013}.

 %\subsection{Hierarchy of Bounds for Multiparameter Estimation}

%\textcolor{magenta}{Possible alternative formulations: "Bounds on the Holevo Bound", "Sandwiching the Holevo Bound", "Upper and Lower Estimates for CH", "Hierarchy of Bounds for Multiparameter Estimation"}
 
 A key challenge in multiparameter estimation arises from the incompatibility of optimal measurements for different parameters. The condition under which the SLD bound is saturable is given by the so-called weak commutativity criterion \cite{ragy2016}:
\begin{equation}
\Tr \left[\rho_{\bm \lambda} [L_\mu^S, L_\nu^S]\right]=0.\label{WCC}
\end{equation}
where $[A,B]=AB-BA$ denotes the commutator of the operators $A$ and $B$.
This motivates the definition of the antisymmetric matrix $U$, often called the mean Uhlmann 
curvature (MUC). This quantity captures the average non-commutativity between the 
parameter generators:
\begin{equation}
{U}_{\mu\nu} \equiv \frac{1}{2i}\Tr \left[ \rho_{\bm \lambda} [L_\mu^S, L_\nu^S]\right]\,,
\end{equation}
and is useful to quantify the incompatibility between parameters.
For pure statistical models,  $\rho_{\bm\lambda} = |\psi_{\bm\lambda}\rangle \langle \psi_{\bm\lambda}|$, we have
\begin{align*}
U_{\mu\nu}  = 4\,\textrm{Im}\big(\ip{\partial_\mu\psi_{\bm\lambda}}{\partial_\nu\psi_\lambda}-\ip{\partial_\mu\psi_{\bm\lambda}}{\psi_{\bm\lambda}}\ip{\psi_{\bm\lambda}}{\partial_\nu\psi_{\bm\lambda}}\big),
\end{align*}
\par
Due to the hardness of finding analytical expressions for the Holevo Bound, two bounds have been introduced. 
These are based on the {\em quantumness} measures $R$  \cite{carollo2019, carollo2020geometry, razavian2020quantumness}, and $T$ \cite{HeWeight,carollo2019,albarelli2020perspective}, defined as
\begin{equation}
    R \equiv ||iQ^{-1} U ||_{\infty},
\end{equation}
\begin{equation}
    T(W) \equiv \frac{\|\sqrt{W} Q^{-1} U Q^{-1} \sqrt{W}\|_1}{\operatorname{Tr}\left[WQ^{-1}\right]},
\end{equation}
where $ ||\cdot||_{\infty}$ denotes the maximum eigenvalue of the matrix, and $||A||_1 = \text{Tr}[\sqrt{A^\dagger A}]$ is the nuclear norm. The corresponding bounds respect the following chain of inequalities:
\begin{equation}
    C_\text{S}[ W,{\hat{\bm \lambda}}] \leq C_\text{H}[ W,{\hat{\bm \lambda}}]\leq 
    \left[1+T(W)\right]\,C_\text{S}[ W,{\hat{\bm \lambda}}]\leq 
    \left[1+R\right]\,C_\text{S}[ W,{\hat{\bm \lambda}}]
    \leq 2\,C_\text{S}[ W,{\hat{\bm \lambda}}]\,.\label{ine}
\end{equation}

\subsection{Two- and Three-Parameter Pure-State Models}\label{sec:2DPureModels}
For the special case of two-parameter pure models, we have that $R$ simplifies to\cite{razavian2020quantumness}
\begin{equation}\label{eq:R2}
	R_2=\sqrt{\frac{\det \left[U\right]}{\det\left[ Q\right] }},
\end{equation}
and using a diagonal weight matrix $W=\text{diag}(1,\omega)$, the $T$ bound \jh{reduces} to
\begin{equation}\label{eq:T2}
    T_2 = 2\, \frac{\sqrt{\omega\,\det [U]}}{Q_{22}+\omega\, Q_{11}}.
\end{equation}
\jh{For two-parameter qubit models we have that \cite{suzuki} 
\begin{equation}
    C_\text{H}[W, \hat{\bm \lambda}] = C_\text{S}[W,  \hat{\bm\lambda}] + \frac{\sqrt{\det [W]}}{\det [Q]}\left|\text{Tr}[\rho_{\bm \lambda}[L_1^S, L_2^S]]\right|.
\end{equation}
For pure two-parameter qubit models we have
\begin{equation}
    \det [Q] = \det [U] =U_{12}^2 = \left(\frac{1}{2i}\text{Tr}\left[\rho [L_1^S, L_2^S]\right]\right)^2\,,
\end{equation}
and thus, using Eq. (\ref{eq:T2}),  the explicit formula
\begin{equation}
C_\text{H} = \Tr[ W Q^{-1}]+ 2\sqrt{\det [ W Q^{-1}]} = C_\text{S}(1+T_2). \label{HT}
\end{equation}
These bounds respect the following bracketing relation
\begin{equation}
    C_\text{S}\leq C_\text{H}= (1+T_2)C_\text{S}\leq (1+R_2)C_\text{S}= 2C_\text{S}\,.\label{ine1}
\end{equation}
}
For three-parameter models $R$ is given by
\begin{equation}
	R_3 = \sqrt{\frac{\textbf{u}^T Q\, \textbf{u}}{\det \left[ Q\right] }},
\end{equation}
with the vector $\textbf{u}$ defined as
\begin{equation}
    \textbf{u}=(U_{23},-U_{13},U_{12}).
\end{equation}
\jh{The quantity $T$, for a diagonal weight matrix $W = \mathrm{diag}(1,\omega_1,\omega_2)$ is given by
\begin{equation}
    T_3 \equiv T_3(W_3)= 2\, \frac{\sqrt{\textbf{u}^{T} Q\,\widetilde{W}_3\, Q\, \textbf{u}}}{\det \left[ Q\right]},
\end{equation}
where 
\begin{equation*}
    \widetilde{W}_3 = \mathrm{diag}(\omega_1\omega_2,\ \omega_2,\ \omega_1).
\end{equation*}
} 
%%%
\section{A Bound for Stepwise Measurement Strategies}
The SLD QCRB offers a computationally straightforward lower bound on the estimator variance, but its attainability is guaranteed only if the SLDs satisfy the weak commutativity criterion.
When this compatibility condition is not met, the Holevo Cramér-Rao Bound provides a tighter, asymptotically achievable limit. However, reaching this bound necessitates collective measurements—experimentally demanding protocols that act jointly on many copies of the quantum state. Nagaoka proposed a simpler measurement strategy using separable measurements performed independently on each probe. While this simplifies the implementation, it generally results in a less tight bound. In this context, we introduce a new class of bounds tailored for sequential measurement strategies. This {\em stepwise} approach involves allocating resources sequentially to estimate subsets of parameters, providing a framework that bridges the gap between single-parameter estimation and fully collective multiparameter strategies.
\par
Consider the task of estimating a set of $n$ parameters, $(\lambda_1,...,\lambda_n)$, 
using a total of $M$ identically prepared probes. \jh{To implement a sequential measurement strategy, we partition the total set of probes into $n$ groups \gf{(each dedicated to a different parameter)} according to allocation fractions $\{\gamma_j:j=1,..,n\},$ where each $\gamma_j$ satisfies $0 \leq \gamma_j \leq 1$ and $\sum_{i=j}^n \gamma_j = 1$. Thus, the $j$-th group contains $\gamma_j M$ probes.} The process unfolds as follows:
\begin{enumerate}
    \item \jh{An experiment with $\gamma_1 M$ probes} is performed, using a measurement strategy optimized to estimate $\lambda_1$ and assuming the other parameters 
    $(\lambda_2,\dots,\lambda_n)$ unknown; 
    \item A second experiment, employing \jh{with $\gamma_2 M$ probes} is performed, aiming at
    optimally estimating $\lambda_2$ assuming $\lambda_1$ known from the previous step and
    the rest of parameters $(\lambda_3,\dots,\lambda_n)$ unknown; 
    \item \jh{The procedure continues sequentially: at step $j$,  $\gamma_j M$ probes are dedicated to optimally estimating} the parameter $\lambda_j$ assuming $\lambda_1,\dots \lambda_{j-1}$ known from the previous steps and
    the rest of parameters $(\lambda_{j+1},\dots,\lambda_n)$ unknown; 
    \item \jh{Finally, the last group of $\gamma_n M$ probes} is used to implement 
    a single-parameter quantum estimation of $\lambda_n$.
\end{enumerate}
The total variance for this protocol is $\sum_j V(\hat{\lambda_j})$, and is bounded by the sum of the optimal variances achievable at each step. By optimizing over all possible resource allocations $\{\gamma_j\}$, we define the stepwise separable bound for a given estimation order $(1\rightarrow n)$ as:
\begin{equation}
    C_{\text{sep}}^{1 \rightarrow n} \equiv \min\limits_{\gamma_1,\dots,\gamma_n} \frac{[Q_{1,\dots,n}^{-1}]_{11}}{\gamma_1} + \frac{[Q_{2,\dots,n}^{-1}]_{11}}{\gamma_2} + ... + \frac{Q_{n}^{-1}}{\gamma_n}\,. \label{csepdef}
\end{equation}
Here, $Q_{j,\dots,n}$ is the Quantum Fisher Information Matrix (QFIM) for the parameter subset $(\lambda_j,\dots,\lambda_n)$, constructed by removing the first $j-1$ rows and columns from the full QFIM. \jh{Its inverse, $Q_{j,\dots,n}^{-1}$, is an $(n-j+1)\times(n-j+1)$ positive-definite matrix. The notation $[Q_{j,\dots,n}^{-1}]_{11}$ refers to the $(1,1)$ entry of this inverse matrix, corresponding to the SLD Cramér–Rao bound for $\lambda_j$ 
assuming $\lambda_1,\dots \lambda_{j-1}$ known from the previous steps and
    the rest of parameters $(\lambda_{j+1},\dots,\lambda_n)$ unknown.} 
    \gf{We use only the first element of the inverse QFIM because this is the variance we get by considering a weight matrix of the form $W=\text{diag}(1,0,\dots)$. We do this since we are interested in estimating just the $j$-th parameter at each step.} The explicit dependence \gf{of the $C_{\text{sep}}$} bound on the estimation sequence order is denoted by the superscript. 

We highlight how this bound is also achievable. We in fact have that for each measurement step we are using a weight matrix $W=\text{diag}(0,\dots,1,\dots,0)$, and with this choice  
the Holevo bound coincides with the respective QCRB. \jh{We also note that this bound is achievable in the asymptotic limit. To see this, consider the sequential protocol itself: at step $j$, we allocate $\gamma_j M$ probes and perform the measurement that is optimal for the reduced parameter set $(\lambda_j,\dots,\lambda_n).$ In the asymptotic limit, using the locally unbiased estimator associated with this measurement,  the variance of $\hat{\lambda_j}$ saturates  $[Q_{j,\dots,n}^{-1}]_{11}/\gamma_1$. Summing over all steps yields the stepwise bound, which is achievable in the asymptotic regime.}
% After performing the optimization over $\{\gamma_i\}$ this bound simplifies to:

% \begin{equation}
%     C_{\textup{sep}}^{1 \rightarrow n} = \left(  \sqrt{\frac{\textup{det}[Q_{2,\dots,n}]}{\textup{det}[Q_{1,\dots,n}]}}  + \sqrt{\frac{\textup{det}[Q_{3,\dots,n}]}{\textup{det}[Q_{2,\dots,n}]}} + \dots +\sqrt{\frac{1}{\textup{det}[Q_{n}]}}\right)^2,
% \end{equation}
% with
% \begin{equation}
%     \gamma_i =  \frac{\sqrt{[Q_{i,\dots,n}^{-1}]_{11}}}{\sum\limits_{j=1}^{n} \sqrt{[Q_{j,\dots,n}^{-1}]_{11}}},
% \end{equation}
% where
% \begin{equation*}
%    [Q_{i,\dots,n}^{-1}]_{11} = \frac{\text{det}[Q_{i+1,\dots,n}]}{\text{det}[Q_{i,\dots,n}]}.
% \end{equation*}

\subsection{Minimization of the Stepwise Bound}
Given an $n$-parameter model, we want to obtain the stepwise bound introduced in the 
previous Section, i.e., to perform the minimization in Eq.(\ref{csepdef}). The following two theorems provide \jh{closed analytical expressions for this minimization. The first theorem expresses the bound as a telescopic-like series, effectively removing the minimization over the $\{\gamma _i\}$ in the original definition and yielding a fully general analytical expression for the stepwise bound. While this formula is general, it involves a sum over $n$ terms, which becomes more and more complex for increasing $n$. The second theorem addresses this issue by exploiting the Cholesky decomposition of the QFIM, providing a computationally efficient alternative for evaluating the bound.} 

\begin{theorem}\label{th:Csep1}
The optimized stepwise measurement bound $C_{\text{sep}}^{1\rightarrow n}$ of 
Eq.(\ref{csepdef}) is given by
\begin{equation}
    C_{\textup{sep}}^{1 \rightarrow n} =\left(  \sqrt{\frac{\textup{det}[Q_{2,\dots,n}]}{\textup{det}[Q_{1,\dots,n}]}}  + \sqrt{\frac{\textup{det}[Q_{3,\dots,n}]}{\textup{det}[Q_{2,\dots,n}]}} + \dots +\sqrt{\frac{1}{\textup{det}[Q_{n}]}}\right)^2,
\end{equation}
which is achieved with
\begin{equation}
    \gamma_j = \frac{\sqrt{[Q_{j,\dots,n}^{-1}]_{11}}}{\sum\limits_{l=1}^{n} \sqrt{[Q_{l,\dots,n}^{-1}]_{11}}}.
\end{equation}
\end{theorem}

\emph{Proof:}
Let us denote $[Q_{j,\dots,n}^{-1}]_{11} \equiv A_j$. We can rewrite the problem as
\begin{equation*}
    C_{\text{sep}}^{1 \rightarrow n} =  \min\limits_{\{\gamma_j\}} \sum\limits_{j=1}^n \frac{A_j}{\gamma_j} = \|\underline{x}\|^2,
\end{equation*}
where \jh{$ \underline{x} \equiv \left(\sqrt{\frac{A_1}{\gamma_1}}, \dots, \sqrt{\frac{A_n}{\gamma_n}}\right)$, and $\|\cdot\|$  denotes the norm of vector.} 
\jh{We define}
\begin{equation*}
    \underline{y} \equiv \left(\sqrt{\gamma_1}, \dots, \sqrt{\gamma_n} \right).
\end{equation*}
Using Cauchy-Schwarz inequality $|\underline{x} \cdot \underline{y}|^2 \leq \|\underline{x}\|^2 \|\underline{y}\|^2$ we get
\begin{equation*}
    \left(\sum\limits_{j=1}^n \sqrt{A_j}\right)^2 \leq \left( \sum\limits_{j=1}^n \frac{A_j}{\gamma_j} \right) \left(\sum\limits_{j=1}^n \gamma_i\right) = \sum\limits_{j=1}^n \frac{A_j}{\gamma_j}.
\end{equation*}
The minimum is thus 
\begin{equation*}
    C_{\text{sep}}^{1 \rightarrow n} =  \min\limits_{\{\gamma_j\}} \sum\limits_{j=1}^n \frac{A_j}{\gamma_j} = \left( \sum\limits_{j=1}^n \sqrt{A_j} \right)^2\,,
\end{equation*}
and is obtained when $\underline{y}$ is parallel to $\underline{x}$, i.e., for
\begin{align*}
 \gamma_j \propto \sqrt{A_j}.
\end{align*}
Normalizing to $\sum_j \gamma_j = 1$, we get the optimal $\{\gamma_j\}$ as
\begin{equation*}
    \gamma_j^* =  \frac{\sqrt{A_j}}{\sum\limits_{l=1}^n \sqrt{A_l}} = \frac{\sqrt{[Q_{j,\dots,n}^{-1}]_{11}}}{\sum\limits_{l=1}^{n} \sqrt{[Q_{l,\dots,n}^{-1}]_{11}}}.
\end{equation*}
To express $ C_{\textup{sep}}^{1 \rightarrow n}$ in terms of determinants, we use the relation
\begin{equation*}
   A_j = [Q_{j,\dots,n}^{-1}]_{11} = \frac{\text{det}[Q_{j+1,\dots,n}]}{\text{det}[Q_{j,\dots,n}]}.
\end{equation*}
With this we prove that
\begin{align*}
    C_{\textup{sep}}^{1 \rightarrow n} = \left( \sum\limits_{j=1}^n \sqrt{A_j} \right)^2 =\left(  \sqrt{\frac{\textup{det}[Q_{2,\dots,n}]}{\textup{det}[Q_{1,\dots,n}]}}  + \sqrt{\frac{\textup{det}[Q_{3,\dots,n}]}{\textup{det}[Q_{2,\dots,n}]}} + \dots +\sqrt{\frac{1}{\textup{det}[Q_{n}]}}\right)^2.
\end{align*}

\begin{corollary}
    The optimal stepwise measurement bound, weighted by a diagonal weight matrix $W = \textup{diag}(\omega_1,\dots,\omega_n)$ is given by
    \begin{align}
        C_{\textup{sep}}^{1\rightarrow n}[ W] &= \min\limits_{\{\gamma_j\}} \frac{\omega_1[Q_{1,\dots,n}^{-1}]_{11}}{\gamma_1} +  \dots +  \frac{\omega_n Q_{n}^{-1}}{\gamma_n}
        = \min\limits_{\{\gamma_j\}} \sum\limits_{j=1}^n \frac{\tilde{A}_j}{\gamma_j} = \left( \sum\limits_{j=1}^n \sqrt{\tilde{A}_j} \right)^2\\
        &= \left(  \sqrt{\frac{\omega_1 \textup{det}[Q_{2,\dots,n}]}{\textup{det}[Q_{1,\dots,n}]}}  + \sqrt{\frac{\omega_2 \textup{det}[Q_{3,\dots,n}]}{\textup{det}[Q_{2,\dots,n}]}} + \dots +\sqrt{\frac{\omega_n}{\textup{det}[Q_{n}]}}\right)^2.
    \end{align}
\end{corollary}

\begin{theorem}\label{th:csep2}
Given $L$, the lower triangular matrix obtained from the Cholesky decomposition of the QFIM $Q=LL^T$, then the stepwise bound $C_{\text{sep}}^{n \rightarrow 1}$ can be expressed as
\begin{equation}
    C_{\text{sep}}^{n \rightarrow 1} = \left(\textup{Tr}[L^{-1}]\right)^2,
\end{equation}
where we highlight that the measurement order is reversed, firstly the $n$-th parameter, up to the first.
\end{theorem}
\emph{Proof:}
Knowing that $L$ is lower triangular, we have that the \gf{right} side is
\begin{equation*}
 \left(\textup{Tr}[L^{-1}]\right)^2 = \left(\sum\limits_{j=1}^n \frac{1}{[L]_{jj}} \right)^2,
\end{equation*}
where $[L]_{jj}$ is the diagonal elements of $L$. For the leading principal minors, we know
\begin{equation*}
    Q_{1,\dots,j} = L_{1,\dots,j}L_{1,\dots,j}^T,
\end{equation*}
 from which we get
\begin{equation*}
    \text{det}[Q_{1,\dots,j}] = \text{det}[L_{1,\dots,j}]\text{det}[L_{1,\dots,j}^T] = \text{det}[L_{1,\dots,j}]^2 = \prod\limits_{l=1}^j [L]_{ll}^2,
\end{equation*}
where $L_{1,\dots,j}$ refers to leading principal minors of order $j$ from $L$.
Since this holds for all $j$, we have
\begin{equation*}
    [L]_{jj} = \sqrt{\frac{\textup{det}[Q_{1,\dots,j}]}{\textup{det}[Q_{1,\dots,j-1}]}},
\end{equation*}
then 
\begin{equation*}
  \textup{Tr}[L^{-1}]= \sum_j \frac{1}{[L]_{jj}} = \sum_i \sqrt{\frac{\textup{det}[Q_{1,\dots,j-1}]}{\textup{det}[Q_{1,\dots,j}]}}.
\end{equation*}
Therefore, we get
% \begin{equation*}
%  \left(\textup{Tr}[L^{-1}]\right)^2=\left(  \sqrt{\frac{\textup{det}[Q_{2,\dots,n}]}{\textup{det}[Q_{1,\dots,n}]}}  + \sqrt{\frac{\textup{det}[Q_{3,\dots,n}]}{\textup{det}[Q_{2,\dots,n}]}} + \dots +\sqrt{\frac{1}{\textup{det}[Q_{n}]}}\right)^2.
% \end{equation*}
\gf{\begin{equation*}
 \left(\textup{Tr}[L^{-1}]\right)^2=\left(  \sqrt{\frac{1}{\textup{det}[Q_{1}]}}  + \sqrt{\frac{\textup{det}[Q_{1}]}{\textup{det}[Q_{1,2}]}} + \dots +\sqrt{\frac{\textup{det}[Q_{1,\dots,n-1}]}{\textup{det}[Q_{1,\dots,n}]}}\right)^2.
\end{equation*}}
\jh{Here we adopt the reversed estimation order, starting from the $n$-th parameter up to the first, in order to simplify the algebraic derivation. The same approach can be applied to obtain a closed-form expression for the original $1 \rightarrow n$ order.}

\subsection{Bounding of the Stepwise Strategy}
The value of the stepwise bound $C_{\text{sep}}$ changes with the order of estimation of the parameters, and a question arises on how these values are related. In Appendix \ref{ap:optimizationCsep} we suggest an algorithm to efficiently calculate numerically 
the more convenient ordering for given a QFIM. In this Section, we instead tackle an order independent analysis. In particular we provide an order independent lower and upper bound.

% We therefore provide a lower and an upper bound for this quantity, independently of the chosen ordering.
\begin{theorem}\label{th:csepBracketing}
For $n$-parameters estimation, the stepwise bound is bracketed by
% \begin{equation}
%  C_{\text{sep}}^{\text{min}}=n^2\sqrt[n]{\frac1{\textup{det} Q}}.
%  \end{equation}
\begin{equation}
 \frac{n^3}{\textup{Tr}[Q]} \leq n^2\sqrt[n]{\frac1{\textup{det} [Q]}} \leq C_{\textup{sep}} \leq n \textup{Tr}[Q^{-1}].
\end{equation}
\end{theorem}

\emph{Proof:}
\jh{We begin by considering the second inequality and, without loss of generality, focus on the stepwise bound} $ C_{\text{sep}}^{1 \rightarrow n}$. Using the 
arithmetic/geometric means (AM-GM) inequality we get
\begin{align*}
    C_{\textup{sep}}^{1 \rightarrow n} &=\left(  \sqrt{\frac{\textup{det}[Q_{2,\dots,n}]}{\textup{det}[Q_{1,\dots,n}]}}  + \sqrt{\frac{\textup{det}[Q_{3,\dots,n}]}{\textup{det}[Q_{2,\dots,n}]}} + \dots +\sqrt{\frac{1}{\textup{det}[Q_{n}]}}\right)^2\\
    &\geq \left(n \sqrt[n]{ \sqrt{\frac{\textup{det}[Q_{2,\dots,n}]}{\textup{det}[Q_{1,\dots,n}]}} \times\sqrt{\frac{\textup{det}[Q_{3,\dots,n}]}{\textup{det}[Q_{2,\dots,n}]}} \times\dots\times\sqrt{\frac{1}{\textup{det}[Q_{n}]}}}\right)^2\\
    &=n^2\sqrt[n]{\frac1{\textup{det} [Q]}}.
\end{align*}
This can then be \gf{reproduced} for any choice of ordering. In turn, changing the order of parameter estimation corresponds to a rearrangement of the rows and columns of the QFIM. Having that $\text{det}[Q]$ is invariant under those changes, this provides a lower limit for all orderings. The  AM-GM also tells that in general this bound is not tight. It is 
achieved iff all the elements of the series are equal, meaning iff all diagonal 
elements of the Cholesky decomposition of the QFIM $L$ are equal.
\par
The leftmost inequality is then proved by a property of positive matrices
\begin{equation*}
    \frac{n}{\Tr A^{-1}}\leq \sqrt[n]{\textup{det} [A]},
\end{equation*}
which can be seen using the harmonic/geometric mean inequality (HM-GM) on the eigenvalues.
\par
The last inequality to prove is the rightmost, and can be seen using that
\begin{equation*}
    [Q_{j,\dots,n}^{-1}]_{11} = \frac{\text{det}[Q_{j+1,\dots,n}]}{\text{det}[Q_{j,\dots,n}]},
\end{equation*}
and
\begin{equation*}
    [Q_{j,\dots,n}^{-1}]_{11} \leq [Q^{-1}]_{jj}.
\end{equation*}
With these, using the arithmetic/quadratic (AM-QM) inequality, we have
\begin{align*}
    C_{\text{sep}} = \left(\sum\limits_{j=1}^n \sqrt{[Q_{j,\dots,n}^{-1}]_{11}} \right)^2 \leq n \sum\limits_{j=1}^n [Q_{j,\dots,n}^{-1}]_{11} \leq n \text{Tr}[Q^{-1}]
\end{align*}
%
% We have the relation:
% \begin{equation}
%  \frac{n^3}{\Tr Q}\leq C_{\text{sep}}^{\text{min}}\leq n\Tr(Q^{-1}).
%  \end{equation}
%
% It is known that changing the order of parameter estimation is reflected in $C_{\textup{sep}}$ as a rearrangement of the relative positions of the elements in the matrix $Q$ corresponding to different parameters. From the above derivation, it can be seen that regardless of how $Q_{i,\dots,n}$ is permuted, these effects cancel out, and only $\textup{det} Q$ remains. Therefore, this bound is independent of the measurement order.
%
% According to the property of positive matrix,  for positive matrix $A\in\mathbb{R}^{n\times n}$, the following inequality holds:
%  $$\frac{n}{\Tr A^{-1}}\leq\sqrt[n]{{\textup{det} A}}\leq\frac{1}{n}\Tr A.$$
% Applying this inequality to the matrix $Q$, we obtain
% \begin{equation*}
%  \frac{n^3}{\Tr Q}\leq n^2\sqrt[n]{\frac1{\textup{det} Q}}\leq n\Tr(Q^{-1}).
%  \end{equation*}
% The necessary and sufficient condition for both the upper and lower bounds to be saturated is $Q=cI_n$, where $c>0$.
%
% According to the property of positive matrix,  for positive matrix $A\in\mathbb{R}^{n\times n}$, the following inequality holds:
%  $$\frac{n}{\Tr A^{-1}}\leq\sqrt[n]{{\textup{det} A}}\leq\frac{1}{n}\Tr A.$$
% Applying this inequality to the matrix $Q$, we obtain
% \begin{equation*}
%  \frac{n^3}{\Tr Q}\leq n^2\sqrt[n]{\frac1{\textup{det} Q}}\leq n\Tr(Q^{-1}).
%  \end{equation*}
A \jh{necessary and} sufficient condition for both the upper and lower bounds to be saturated is $Q=cI_n$, where $c>0$.
\jh{If $Q=cI_n$, we easily obtain lower and upper bounds coincide with $C_{\text{sep}}$. Conversely, suppose $C_{\text{sep}}$ simultaneously reaches the lower and upper bounds. From the lower bound (AM–GM), all terms in the telescopic product must be equal: $[Q_{j,\dots,n}^{-1}]_{11}=k^2$ for all $j$. From the upper bound (AM–QM), we must have $ [Q_{j,\dots,n}^{-1}]_{11} = [Q^{-1}]_{jj}$ and all diagonal elements of $Q^{-1}$ equal. Combining these conditions with the fact that  $Q$ is a positive definite symmetric matrix, implies $Q=\frac1{k^2}\,I_n$. Thus, $Q=c\,I_n$ is both necessary and sufficient.}

\subsection{$C_{\text{sep}}$ and sloppiness in 2-parameter qubit models}\label{sec:2Dmodels}
In the case of a two-parameter qubit model, we now establish conditions for the separable $C_{\text{sep}}$ to outperform the three bounds: SLD QCRB $ C_\text{S}$, $T$-dependent bound $C_\text{T}$, and $R$-dependent bound $C_\text{R}$. %Crucially, $C_{\text{sep}}$'s physical attainability depends on the estimation order, distinguishing it from symmetric bounds.
\gf{Since $C_\text{sep}$ is order dependent,} we focus on $C_{\text{sep}}^{1\rightarrow2}$ (estimating $\lambda_1$ before $\lambda_2$), with formulas for the opposite order derivable by switching $Q_{11}$ and $Q_{22}$. To isolate QFIM effects, we set $W = \text{diag}(1,1)$ in $C_\text{S}$ and $C_\text{T}$.

To understand when $C_{\text{sep}}^{1\rightarrow2} \leq C_\text{S}$ holds, we convert the inequality into a condition on the amount of sloppiness, which is quantified by the parameter $s$ \cite{sloppiness}
$$s := \frac{1}{\det[Q]},$$
 which captures how strongly the system depends on a combination of the parameters
 rather than on them individually.
 In particular, when estimating $\lambda_1$ first, followed by the estimation of 
 $\lambda_2$, the separable bound is given by 
 $$C_{\text{sep}}^{1\rightarrow2}=\frac{1}{Q_{22}}+\frac{Q_{22}}{{\text{det} [Q]}}+2\sqrt{\frac{1}{{\text{det} [Q]}}}\,,$$
 and the inequality $C_{\text{sep}}^{1\rightarrow2} \leq C_\text{S}$ may be
 written as
\begin{equation}\label{conditionCS}
s \geq \frac{1}{Q_{11}^2} \left(1+\sqrt{1+ Q_{11}/Q_{22}}\right)^2\equiv \frac{4 Q_{22}^2}{Q_{12}^4}\,.
\end{equation}

 This indicates that $C_{\text{sep}}^{1\rightarrow2}$ becomes tighter if sloppiness is sufficiently large. The threshold is determined by the ratio $Q_{22}/Q_{12}^2$, not by their absolute values. When $|Q_{12}| \gg \sqrt{Q_{22}}$ (strong correlation), $C_{\text{sep}}$ outperforms $C_\text{S}$ even with moderate sloppiness. However, from the above expression, it is clear that when the QFIM is diagonal (i.e., $Q_{12} = 0$), $C_{\text{sep}}^{1\rightarrow2}$ is strictly larger than $C_\text{S}$. This is consistent with the intuition that when parameters are statistically independent (i.e., the QFIM is diagonal) stepwise estimation offers no advantage. Conversely, when statistical correlations between parameters are unavoidable due to model structure, a stepwise estimation strategy may become advantageous.
 
\gf{Since} $C_\text{S}$ is not always attainable, we now discuss the relationships of 
$C_{\text{sep}}$ with  the two bounds $C_\text{T}$ and $C_\text{R}$, which are analytically expressible %and in some 2-parameter estimation cases attainable
\gf{and always saturable (at least asymptotically) and provide upper bounds 
for $C_\text{H}$} at fixed weight matrix and for all of them, respectively. 
The bound
$C_\text{T}$ is  given by
\begin{equation*}
C_\text{T} = C_\text{S}(1+T_2)=\frac{Q_{11} + Q_{22} + 2\abs{U_{12}}}{\text{det}[Q]},
\end{equation*}
and the condition
\jh{$C_{\text{sep}}^{1\rightarrow2} \leq C_\text{T}$} 
corresponds to
\begin{equation}
s\geq \frac{4Q_{22}^2}{\left(Q_{12}^2 + 2Q_{22}\abs{U_{12}}\right)^2}.
\end{equation}
This implies that $C_{\text{sep}}$ surpasses $C_\text{T}$ under weaker sloppiness requirements, enhancing its practical attainability. Furthermore, when the QFIM is diagonal, the condition for $C_{\text{sep}}^{1\rightarrow2} \leq C_\text{T}$ becomes $s \geq \frac{1}{\text{det} [U]}$, which is always satisfied in two-parameter pure-state qubit models according to \cite{sloppiness}. Furthermore, this implies that in such models: $C_{\text{sep}}^{1\rightarrow2} = C_\text{T} = C_\text{H}$, and the advantage of $C_{\text{sep}}^{1\rightarrow2}$ becomes evident when sloppiness is large enough.

The $R$-dependent bound reads as follows
\begin{equation*}
C_\text{R} =C_\text{S}(1+R_2)= \frac{Q_{11} + Q_{22}}{\detQ} \left(1 + \abs{U_{12}} \sqrt{\frac{1}{\text{det}[Q]}} \right),
\end{equation*}
and the inequality $C_{\text{sep}}^{1\rightarrow2} \leq C_\text{R}$ corresponds to
\begin{align}\label{eq:CR_inequality} 
\abs{U_{12}}(Q_{12}^2 + Q_{22}^2)s+Q_{12}^2 \sqrt{s} +\abs{U_{12}}-2Q_{22}\geq0.
\end{align}
Solving \eqref{eq:CR_inequality} yields the explicit condition:
\begin{equation*}
s \geq \frac{\left( \sqrt{\Delta} - Q_{12}^2 \right)^2}{4\abs{U_{12}}^2 \left(Q_{12}^2 + Q_{22}^2\right)^2}, \quad \Delta = Q_{12}^4 + 4\abs{U_{12}} \left(Q_{12}^2 + Q_{22}^2\right)(2 Q_{22} - \abs{U_{12}})\geq0.
\end{equation*}
Remarkably, when $2 Q_{22} = \abs{U_{12}}$, the condition simplifies to $s \geq 0$, indicating that $C_{\text{sep}}^{1\rightarrow2}$ universally outperforms $C_\text{R}$ regardless of the sloppiness. This scenario corresponds to a regime where quantum incompatibility makes sequential estimation mostly convenient. When the QFIM is diagonal, the condition simplifies significantly, and the threshold becomes
$$
s \geq \dfrac{2Q_{22}-|U_{12}|}{|U_{12}|Q_{22}^2} .
$$
When $|U_{12}| > 2 Q_{22}$,  the inequality holds for all $s > 0$ since the right-hand side is negative, and so $C_{\text{sep}}^{1\rightarrow2}$ always dominates $C_\text{R}$.  When $|U_{12}| < 2 Q_{22}$, the proper relation between $Q_{22}$ and $\abs{U_{12}}$ allows $C_{\text{sep}}^{1\rightarrow2} \leq C_\text{R}$ to hold even with low sloppiness, e.g., when $Q_{22}$ and $\abs{U_{12}}$ are both large.
In the special case of vanishing incompatibility (i.e., $U_{12} = 0$), the attainability condition for $C_\text{T}$ and $C_\text{R}$ reduces exactly to that of $C_{\text{sep}}^{1\rightarrow2}$ as shown in Eq. (\ref{conditionCS}). However, even in these cases, the inequality only holds when sloppiness is sufficiently large. Finally, while the expression for $C_{\text{sep}}^{1\rightarrow2}\leq C_\text{R}$ is more involved, it still ultimately requires a large sloppiness for the superiority of $C_{\text{sep}}$ to manifest.

In summary, $C_{\text{sep}}$ offers a practically attainable bound in models with sufficient sloppiness, especially when the interplay between quantum incompatibility and statistical correlation is optimized.

\section{Performance of joint and stepwise estimation in SU(2) models}
In this Section, we compare the performance of joint estimation (JE) and stepwise estimation (SE) strategies in estimating the parameters encoded unitarily by $SU(2)$ Hamiltonians of 
the form 
\begin{equation}
    H = \mathbf{B} \cdot \textbf{J}_{\mathbf{n}},
\end{equation}
where $\mathbf{J}_{\textbf{n}}$ is an $SU(2)$ generator along direction $\textbf{n}$. Three cases are considered: a two-parameter model with qubits, a two-parameter model with qutrits, and a three-parameter model with qutrits. Here, we summarize the main results and discuss their implications for multiparameter quantum estimation. The detailed derivations of the QFIM and the Uhlmann matrix for each case are provided in Appendix \ref{ap:twoparams}. 

 \subsection{SU(2) 2-parameter estimation in qubit models }\label{sec:SU22parsqubit}
Let us consider the Hamiltonian
\begin{equation}
    H_{B,\theta} = B(\cos \theta J_x + \sin \theta J_z)\,,
\end{equation}
where $J_j$ denotes the $j$-th generator of $SU(2)$. As discussed in \cite{dimensionMatters}, in the case of qubit probes the QFIM is
\begin{align}
    &Q_{BB} = t^2 [1-(\textbf{n}_\theta \cdot \textbf{r}_0)^2]\\
    &Q_{\theta\theta} = 4 \sin^2 \frac{Bt}{2} [1-(\textbf{n}_1\cdot\textbf{r}_0)^2]\\
    &Q_{B\theta} = 2t\sin\frac{Bt}{2}(\textbf{n}_1 \cdot \textbf{r}_0)(\textbf{n}_\theta \cdot \textbf{r}_0),
\end{align}
and the Uhlmann matrix is
\begin{equation}
    D_{\theta B} = 2t\sin\frac{Bt}{2}\textbf{n}_2\cdot\textbf{r}_0
\end{equation}
with vectors
\begin{align}
    &\textbf{n}_\theta = (\cos \theta, 0, \sin \theta)\\
    &\textbf{n}_1 = (\cos\frac{Bt}{2}\sin\theta,-\sin\frac{Bt}{2},-\cos\frac{Bt}{2}\cos\theta)\\
    &\textbf{n}_2 = \textbf{n}_\theta \times \textbf{n}_1 = (\sin\frac{Bt}{2}\sin\theta,\cos\frac{Bt}{2},-\sin\frac{Bt}{2}\cos\theta)\\
    &\textbf{r}_0 = (\text{Tr}[\sigma_x \rho_0],\text{Tr}[\sigma_y \rho_0],\text{Tr}[\sigma_z \rho_0]),
\end{align}
where $\rho_0 = \ket{\psi_0}\bra{\psi_0}$ denotes the input probe state.
\par
\jh{Our goal is to compare the performance of stepwise and joint estimation strategies by analyzing the SE bound alongside other bounds, including the Holevo bound. Since the SLD Cramér–Rao bound is not attainable in this model, our analysis focuses on the Holevo bound, together with the $T$-bound and the $R$-bound.}
For these kind of qubit models, we always have $R=1$ \cite{dimensionMatters,razavian2020quantumness}, independently on the value 
of the parameters $(B,\theta)$ and the choice of the probe state $\ket{\psi_0}$. 
We thus have $C_\text{R} = 2 C_\text{S}$. From Theorem \ref{th:csepBracketing} we also have $C_{\text{sep}} \leq 2\,\text{Tr}[Q^{-1}]\equiv 2 C_\text{S}$ for two-parameter models. Combining these results immediately gives
\begin{equation}
    C_{\text{sep}} \leq C_\text{R}
\end{equation}

Next, we consider the $T$-bound. Using Eq. \ref{eq:T2} with the identity as weight matrix, 
we get
\begin{equation}
    T = \frac{4 t \sqrt{(1-\alpha ^2-\beta ^2)\sin^2\frac{Bt}{2}}}{2 \left(1-\beta ^2\right) (1-\cos (B t))+\left(1-\alpha ^2\right) t^2},
\end{equation}
where $\alpha \equiv \textbf{n}_\theta\cdot\textbf{r}_0$ and $\beta \equiv \textbf{n}_1\cdot\textbf{r}_0$, such that $\alpha,\beta \in [-1,1]$, $\alpha^2+\beta^2\leq 1$ and $\textbf{n}_2\cdot\textbf{r}_0 = \sqrt{1-\alpha^2-\beta^2}$. From this, it can be proven analytically that
\begin{equation}
    C_{\text{sep}}^{1\rightarrow 2},C_{\text{sep}}^{2\rightarrow 1} \leq \text{Tr}[Q^{-1}](1+T)
\end{equation}
holds for every choice of $\alpha,\beta,B,t$ whenever $Q$ is invertible. 
As stated in Eq. (\ref{HT}) for this model $C_\text{H} = C_\text{S}(1+T). $
Therefore, we conclude that for two-parameter  $SU(2)$ qubit models a stepwise estimation strategy always outperform the joint strategy:
\begin{equation}
 C_{\text{sep}}^{1\rightarrow 2},C_{\text{sep}}^{2\rightarrow 1} \leq C_\text{T} =C_\text{H}.
\end{equation}

%A detailed comparison between the SE bound and the Cramér–Rao bound is provided in Appendix~\ref{ap:sevsCRB}\jh{remove this appendix?} \gf{I agree}.

\subsection{SU(2) \jh{2-parameter estimation in qutrit models }}
%To tackle this we approach the problem in a similar way as we did for qubits, using vectors. We consider as a base the Gell-Mann matrices (normalized) plus the identity $\textbf{b} = \{\lambda_1/\sqrt{2}, ..., \lambda_8/\sqrt{2}, \mathbb{I}/\sqrt{3}\}$, so that $\textbf{b}_i\cdot\textbf{b}_j = \delta_{ij}$. Then, the $J_\textbf{n}$ matrices can be built with the Pauli spin 1 matrices
\jh{ We now consider the a two-parameter $SU(2)$ qutrit model, where the Hamiltonian remains the same as in the qubit case, but the  probe state is generalized to a qutrit. The detailed calculations are provided in Appendix \ref{ap:twoparams}.}

% \jh{(I change the structure of this part and put some backgrounds to Appendix.) }
%This way, we can rewrite expectation values as
%\begin{align}
%   \braket{A}_0 &= \text{Tr}[A \rho_0] = \text{Tr}[(\sum_i \textbf{A}_i \textbf{b}_i)(\sum_j \textbf{r}_j \textbf{b}_j)] = \sum_i \sum_j \textbf{A}_i \textbf{r}_j \text{Tr}[\textbf{b}_i \textbf{b}_j]\\
 %  &= \sum_i \textbf{A}_i \textbf{r}_i = \textbf{A}\cdot\textbf{r},
%\end{align}
%with $\textbf{A}_i = \text{Tr}[A \textbf{b}_i]$ and $\textbf{r}_i = \text{Tr}[\rho \textbf{b}_i]$.

% We obtain the QFIM
\gf{The QFIM is}
\begin{align}\label{eq:1Qsu2qutrit} 
   &Q_{BB} = 4 t^2 [\textbf{n}_{\theta}^2 \cdot \textbf{r}-(\textbf{n}_{\theta}\cdot\textbf{r})^2]\\
   \label{eq:2Qsu2qutrit}
   &Q_{\theta\theta} = 16 \sin^2 \frac{Bt}{2} [\textbf{n}_1^2\cdot\textbf{r}-(\textbf{n}_1\cdot\textbf{r})^2]\\
   \label{eq:3Qsu2qutrit}
   &Q_{B\theta} = -4t\sin\frac{Bt}{2}[\textbf{n}_{\{1,\theta\}}\cdot\textbf{r}-2(\textbf{n}_1\cdot\textbf{r})(\textbf{n}_\theta\cdot\textbf{r})],
\end{align}
with
% $$
% \textbf{n}_{\theta,i}^2 \equiv \mathrm{Tr}[J_{\textbf{n}_{\theta}}^2 b_i], \quad
% \textbf{n}_{1,i}^2 \equiv \mathrm{Tr}[J_{\textbf{n}_{1}}^2 b_i], \quad
% \textbf{n}_{\{1,\theta\},i} \equiv \mathrm{Tr}[\{J_{\textbf{n}_{1}}, J_{\textbf{n}_{\theta}}\} b_i],
% $$
$$
\gf{[\textbf{n}_{\theta}^2]_i \equiv \mathrm{Tr}[J_{\textbf{n}_{\theta}}^2 b_i], \quad
[\textbf{n}_{1}^2]_i \equiv \mathrm{Tr}[J_{\textbf{n}_{1}}^2 b_i], \quad
[\textbf{n}_{\{1,\theta\}}]_i \equiv \mathrm{Tr}[\{J_{\textbf{n}_{1}}, J_{\textbf{n}_{\theta}}\} b_i],}
$$
and
$$
\{b_i\} = \left\{\frac{\Gamma_1}{\sqrt{2}}, \dots, \frac{\Gamma_8}{\sqrt{2}}, \frac{\mathbb{I}}{\sqrt{3}}\right\}, \qquad
\textbf{r} = \{r_i\} = \big(\mathrm{Tr}[\rho_0 b_1], \dots, \mathrm{Tr}[\rho_0 b_9]\big),
$$
\gf{with $\Gamma_i$ the Gell-Mann matrices, and $b_i$ their normalized version as described in Appendix \ref{ap:twoparams}.}
The difficulty in optimizing this expression lies in the fact that pure qutrit states live on a 4-dimensional manifold embedded in a 7-sphere, making the domain highly nontrivial. To address this, we resort to numerical optimization, sampling over all possible pure qutrit states. Our results reveal that the optimal states for various values of $B, \theta,$ and $t$ share a common property, they satisfy
$$
\textbf{r}\perp \{\textbf{n}_\theta, \textbf{n}_1, \textbf{n}_{\{1,\theta\}}, \textbf{n}_2\},
\quad \text{and} \quad
\textbf{n}_\theta^2 \cdot \textbf{r}, \, \textbf{n}_1^2 \cdot \textbf{r} = 1.
$$
This implies that the QFIM takes the form
\begin{equation}
    \label{eq:nondiagW}
    Q = \left(\begin{array}{cc}
 4 t^2 & 0 \\
 0 & 16 \sin^2 \frac{Bt}{2} \\\end{array}\right),
\end{equation}
yielding a \gf{QCRB} of
$$C_\text{S}=\frac{1}{16}\left(\frac{4}{t^2}+\csc^2\frac{Bt}{2}\right).$$ The Uhlmann curvature becomes
\begin{equation}
    \label{eq:UnondiagW}
    U = \left(\begin{array}{cc}
 0 & -4t \sin \frac{Bt}{2} (\textbf{n}_2 \cdot \textbf{r}) \\
 4t \sin \frac{Bt}{2} (\textbf{n}_2 \cdot \textbf{r}) & 0 \\\end{array}\right) = \mathbf{0},
\end{equation}
since $\textbf{r} \perp \textbf{n}_2$.
\par
The above analysis shows that using qutrits we have sufficient freedom (i.e., room for improvement in the Hilbert space) to minimize the quantum incompatibility. The Uhlmann matrix vanishes, $U=\mathbf{0}$, and thus we have $T=0$ and $C_\text{H} = C_{\text{SLD}}$. Furthermore, the optimal QFIM is diagonal. We conclude that by suitably choose the probe, joint estimation (JE) outperforms  stepwise estimation (SE).
%%%
\subsection{SU(2) \jh{3-parameter estimation in qutrit models }}
In the previous two examples, we analyzed two-parameter models and observed that increasing the probe dimension reduces the effectiveness of the SE method when the encoding is good. We now explore how this behavior changes when the number of parameters increases to three. If we restrict to qubits, the small dimension of the Hilbert space makes the QFIM singular. Therefore, the smallest probe dimension that allows a meaningful analysis is a qutrit. We extend the previous model by allowing the vector $\mathbf{B}$ to span all directions. The Hamiltonian is then given by
\begin{equation}
    H_{B,\theta,\phi} = B(\cos \theta \cos \phi J_x + \cos \theta \sin \phi  J_y + \sin \theta  J_z) \,,
\end{equation}
with
\begin{equation}\label{eq:n_theta^3}
    \textbf{n}_\theta^{(3)} = \left(\cos \theta \cos \phi, \cos \theta \sin \phi, \sin \theta\right).
\end{equation}
For more details on the derivation see \cite{dimensionMatter}. Since this model is an extension of the previous one, the terms $Q_{BB}$, $Q_{\theta\theta}$, and $Q_{B\theta}$ remain formally the same as in Eqs. (\ref{eq:1Qsu2qutrit})–(\ref{eq:3Qsu2qutrit}), but now evaluated using the vectors in Eq. (\ref{eq:n_theta^3}) and
\begin{align}
\begin{split}
    \textbf{n}_{1}^{(3)} = \ &\Big(\sin\frac{Bt}{2}\sin\phi + \cos\frac{Bt}{2}\sin\theta\cos\phi,\\
    &-\sin\frac{Bt}{2}\cos\phi + \cos\frac{Bt}{2}\sin\theta\sin\phi, -\cos\frac{Bt}{2}\cos\theta \Big).
\end{split}
\end{align}
In addition, we now have the third parameter $\phi$, leading to the extra QFIM components
\begin{align}
    &Q_{\phi\phi} = 16 \sin^2 \frac{Bt}{2} \cos^2 \theta [\braket{J_{\textbf{n}_{2}^{(3)}}^2}_0-\braket{J_{\textbf{n}_{2}^{(3)}}}_0^2]\\
    &Q_{B\phi} = -4t\sin\frac{Bt}{2} \cos \theta[\braket{\{J_{\textbf{n}_{2}^{(3)}},J_{\textbf{n}_{\theta}^{(3)}}\}}_0-2\braket{J_{\textbf{n}_{2}^{(3)}}}_0 \braket{J_{\textbf{n}_{\theta}^{(3)}}}_0]\\
    &Q_{\theta\phi} = 8\sin^2\frac{Bt}{2} \cos \theta[\braket{\{J_{\textbf{n}_{1}^{(3)}},J_{\textbf{n}_{2}^{(3)}}\}}_0-2\braket{J_{\textbf{n}_{1}^{(3)}}}_0 \braket{J_{\textbf{n}_{2}^{(3)}}}_0], 
\end{align}
with% \underline{(note: this is different from the one in DimensionMatter)}
% \begin{equation}
%     \textbf{n}_{2}^{(3)} = \frac{1}{2}\cos\theta\csc\frac{Bt}{2}\left(
% \begin{array}{c}
%  (\cos (B)-1) \sin \theta  \cos \phi +\sin B \sin \phi  \\
%  (\cos (B)-1) \sin \theta  \sin \phi +\sin B \cos \phi  \\
%  -(\cos B-1) \cos \theta  \\
% \end{array}
% \right).
% \end{equation}
\begin{align}
\begin{split}
    \textbf{n}_2^{(3)} = \ &\Big( \cos\frac{Bt}{2}\sin\phi - \sin\frac{Bt}{2} \sin\theta \cos\phi,\\
    &-\cos\frac{Bt}{2} \cos\phi - \sin\frac{Bt}{2} \sin\theta \sin\phi, \sin\frac{Bt}{2} \cos\theta \Big).
\end{split}
\end{align}

\begin{figure}[h]
    \centering
    \includegraphics[width=1\linewidth]{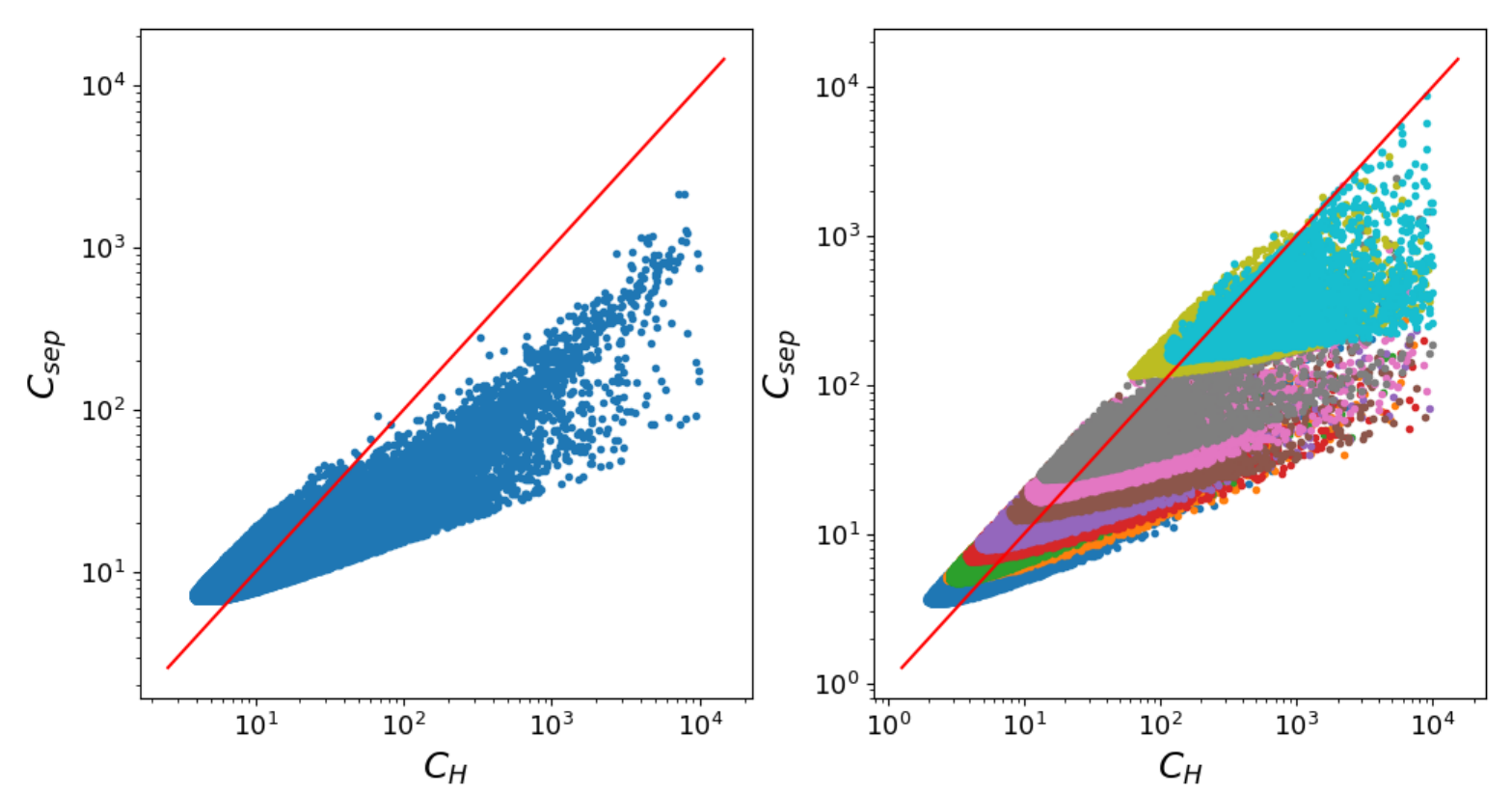}%{chcsepstates.png}
    \caption{Comparison of stepwise and joint estimation for 3-parameter $SU(2)$ qutrit models. In the left panel, we show the results 
for a sample of $10^5$ states and a fixed set of parameters $(B,\theta,\phi)$. In the right one, we consider $10$ different sets of values for $(B,\theta,\phi)$ and sample  $10^4$ states for each set. In both panels, the red line represents the points where $C_\text{H} = C_{\text{sep}}$.}
    \label{fig:chcsepstates}
\end{figure}

Given the complexity of the model, a numerical approach has been considered. Our goal is to investigate when $C_{\text{sep}}$ is smaller than the Holevo bound and how this behavior depends on the characteristics of the probe state. To this end, we sample %$100{,}000$
a large number of random states \gf{given} a fixed set of parameters $(B,\theta,\phi)$. For each probe state, we compute the minimum $C_{\text{sep}}$ over all possible orderings and the Holevo bound (via the SDP approach in \cite{albarelli2019evaluating,genoni25}). The results are shown in Fig. \ref{fig:chcsepstates}. \gf{In the left panel, we show the results 
for a sample of $10^5$ states and a fixed set of parameters $(B,\theta,\phi)$. In the right one, we consider $10$ different sets of values for $(B,\theta,\phi)$ and sample  $10^4$ states for each set. Notice that the choice of different parameter sets $(B,\theta,\phi)$ does not alter the shape of the distribution} . %For clarity, states are ordered by increasing $C_\text{H}$ values.
The red line represents the points where $C_\text{H} = C_{\text{sep}}$\gf{, which separates the two regions where one of the two is larger than the other}.
%\gf{Additionally, we can see that this shape is shared across different parameter choices. In Fig. \ref{fig:chcsepparams} we repeat this graph for 9 choices of $(B,\theta,\phi)$.}
\jh{Although it is evident that for small values of $C_\text{H}$ the Holevo bound is generally tighter ($C_\text{H} < C_{\text{sep}}$), a significant number of states exhibit the opposite behavior, where $C_{\text{sep}} < C_\text{H}$. This observation indicates that, within the three-parameter qutrit estimation model, the Holevo bound is not always the ultimate lower bound. In certain specific cases, $C_{\text{sep}}$ can serve as a tighter or alternative bound for parameter estimation.}

%Nowm we focus on the state achieving the smallest $C_\text{H}$ (on the left-most point in Fig.\ref{fig:chcsepstates}). We repeat this analysis for $10$ different parameter choices $(B,\theta,\phi)$, each time sampling $10{,}000$ states. From each case, we extract the minimum $C_\text{H}$ and the corresponding $C_{\text{sep}}$, and report these in Fig. \ref{fig:chcsepparams}.

% In addition to this, to properly answer the first question we're going to focus on the left-most point, the one with minimum $C_\text{H}$ (and the relative value of $C_{\text{sep}}$. \jh{What does “focus on the left-most point” mean? Does it refer to selecting the states concentrated in this region?
% } We're going to consider 10(??change) of these plots (each with a different $(B,\theta,\phi)$ choice), and take from them just these minimum values, putting them in a single plot. This way we get Fig. \ref{fig:chcsepparams}.

% \begin{figure}[h]
%     \centering
%     \includegraphics[width=0.8\linewidth]{chcsepparams.png}
%     \caption{Using the 3-parameter model of $SU(2)$ with qutrits, we consider \gf{9} random parameter choices $(B,\theta,\phi)$. We then stack over graphs in the form of Fig. \ref{fig:chcsepstates}.}
%     \label{fig:chcsepparams}
% \end{figure}

\gf{We conclude that also in this case, when we have enough quantum resources (i.e., the optimal, or nearly optimal probe) JE is convenient. If the probe state is sub-optimal SE progressively outperforms JE.}

% Looking at both pictures we can now answer to the two questions. The first has a negative answer, since we can see from Fig. \ref{fig:chcsepparams} that the Holevo bound is always smaller than the SE one for the optimal probes. But, looking at Fig. \ref{fig:chcsepstates} we have that the SE bound outperforms the joint schema when the encoding becomes progressively worse.

% \jh{In Fig. \ref{fig:chcsepparams} , we choose the different values of the parameters to be estimated. The regions formed by different colors in the figure correspond to different parameter pairs. It is observed that for certain parameter ranges, there are still many cases where $C_{\text{sep}}$ achieves a tighter bound than $C_\text{H}$. In other words, if the system and parameters to be estimated cannot be adjusted or optimized, $C_{\text{sep}}$ may still outperform $C_\text{H}$ for particular states and parameter ranges.
% }

\section{Conclusions}
In this work, we have analyzed stepwise estimation strategies for multiparameter quantum metrology, deriving a tight and analytically tractable precision bound, the stepwise separable bound $C_\text{sep}$, 
that depends explicitly on the order of parameter estimation. We provided closed-form expressions 
for this bound and showed how it can be efficiently computed using the Cholesky decomposition of the quantum Fisher information matrix (QFIM). We also established rigorous bounds on  $C_\text{sep}$, that hold independently of the estimation order.

Through the analysis of SU(2) unitary models with qubit and qutrit probes, we demonstrated that stepwise estimation can outperform joint estimation strategies, particularly in scenarios characterized by large sloppiness, non-optimal probe states, or strong parameter incompatibility. In two-parameter qubit models,  $C_\text{sep}$ was shown to be tighter than the Holevo bound. For qutrit systems, however, the increased dimensionality allows for better suppression of incompatibility, making joint estimation more favorable under optimal conditions. In three-parameter qutrit models, we identified regimes where  $C_\text{sep}$ remains competitive with or even superior to the Holevo bound, especially for suboptimal encodings. These results establish stepwise estimation as a viable and experimentally friendly alternative to collective measurements, particularly in resource-constrained or imperfect settings where ideal probe states or joint measurements are not feasible.

Our results pave the way for extension to more general quantum systems, including continuous-variable and open quantum systems, and explore adaptive stepwise protocols that dynamically optimize the estimation order and resource allocation. Additionally, the connection between sloppiness, incompatibility, and estimation efficiency warrants further experimental investigation, potentially leading to new design principles for quantum sensors \cite{bizzarri2025controlling}.

\appendix

\section{Dynamical Programming Algorithm for Best Ordering}\label{ap:optimizationCsep}
%\jh{I don't know much about algorithm complexity. But I recommend to put this process in appendix for those are interested in algorithmic complexity details. Readers care about the main theoretical contribution. For example, why the order matters or how to get the best order.}
In Theorem. \ref{th:Csep1} we provided a way to calculate the $C_{\text{sep}}$ bound for a given ordering. As mentioned the result depends on the ordering chosen. In this \gf{Appendix} we want to discuss the \gf{computational} complexity of \gf{finding the optimal ordering that minimizes the bound}, and \gf{to} provide a more efficient way to calculate it through a dynamical programming (DP) approach.\\

Let's start by analyzing the brute force approach. We have to compute all possible orderings of $n$ numbers, therefore $\mathcal{O}(n!)$. For each ordering, we have to compute the trailing determinants. The complexity of computing the determinant of a generic $k\times k$ matrix using methods like LU decomposition is $\mathcal{O}(k^3)$. Considering all the trailing submatrices we therefore have the cost $\sum_{k=1}^{n} \mathcal{O}(k^3) = \mathcal{O}(n^4)$. Therefore, the brute force time complexity is $\mathcal{O}(n!n^4)$.
\par
We can do better than the brute force approach. Noticing a similarity with the traveling salesman problem, we can write a variation on the Bellman-Held-Karp algorithm \cite{bellman1962,heldkarp1962}. The core principle is that given a set of indices $S = \{i_1,\dots,i_k\}$, the optimal sequence must contain an optimal subsequence of $k-1$ elements in the sequence. This property is known as optimal substructure. This allows us to create solutions iteratively, starting from an empty set. \gf{Wanting to find the minimum $C_\text{sep}$, we refer to \ref{th:csep2} and choose the cost function as $\mathcal{C}([1,\dots,n])\equiv \text{Tr}[L^{-1}]$ without the squared part in order to simplify the algebra.} The algorithm for an $n$-paramter model works as follows:
\begin{enumerate}
    \item Base case: the cost for an empty set of parameters is 0, $\mathcal{C}(\emptyset) = 0$.
    \item Recurrence relation: exploiting the optimal substructure, for any non-empty sequence $S$, the optimal cost $\mathcal{C}(S)$ is found by considering each element $j \in S$ as the potential last element in the optimal sequence for $S$. If $j$ is the last element, the preceding elements must form an optimal sequence for the subset $S \setminus \{j\}$. The total cost is thus the optimal cost of the preceding subset plus the cost contribution of adding $j$. We therefore have
\begin{equation}
    \mathcal{C}(S) = \min\limits_{j \in S} \ \mathcal{C}(S \setminus \{j\}) + \text{cost}(j, S\setminus \{j\}).
\end{equation}
\end{enumerate}
The cost function $\text{cost}(j, S\setminus \{j\})$, is that of adding the contribution of the parameter $j$, at the end of the sequence $I \equiv S\setminus \{j\}$. Remembering from Th. \ref{th:csep2} that using $L$ gives a revers ordering, we underline how this means we're measuring the $j$-th parameter before the reversed $I$ sequence. To evaluate the cost addition we consider the new QFIM
\begin{equation}\label{eq:qfimextended}
    Q_S = \left(\begin{array}{cc}
Q_{I,I} & q_{I,j}\\
q_{j,I} & q_{j,j}\\
\end{array}\right),
\end{equation}
with its new Cholesky decomposition matrix
\begin{equation}
    L_S = \left(\begin{array}{cc}
L_{I,I} & 0\\
l_{j,I} & l_{j,j}\\
\end{array}\right).
\end{equation}
From this we see
\begin{equation}
    Q_S = L_S L_S^T = \left(\begin{array}{cc}
L_{I,I}L_{I,I}^T & L_{I,I}l_{j,I}^T\\
l_{j,I}L_{I,I}^T & l_{j,j}^2+l_{j,I}l_{j,I}^T\\
\end{array}\right).
\end{equation}
Equating this to Eq. \ref{eq:qfimextended} we get
\begin{align}
    l_{j,I} = q_{j,I}\left[L_{I,I}^T\right]^{-1} \ \Rightarrow \ l_{j,j}^2 = q_{j,j} - q_{j,I}Q_{I,I}^{-1} q_{j,I}^T,
\end{align}
which is the Schur complement of the block $Q_{I,I}$ of the new QFIM $Q_S$. Using Th. \ref{th:csep2} \gf{on the new sequence we know} that the cost function increases by $1/l_{j,j}$, therefore we conclude
\begin{equation}
    \text{cost}(j, S\setminus \{j\}) = \frac{1}{\sqrt{q_{j,j} - q_{j,I}Q_{I,I}^{-1} q_{j,I}^T}}.
\end{equation}
The algorithm is implemented by iteratively computing $\mathcal{C}(S)$ for all subsets $S$ of size from $0$ to $n$. We use memoization, storing all optimal cost values of subsequences $I$, which will be used for next steps, avoiding redundant calculations. The optimal sequence can then be reconstructed from the intermediate choices (and then reversing them), and the $C_{\text{sep}}$ bound is obtained squaring the optimal value \gf{since} the algorithm minimizes the quantity $\mathcal{C} = \sum_{i=1}^n 1/L_{ii}$.
\par
Let's analyze the complexity of this algorithm. We still have to go through all possible subsets, therefore there's a complexity of $\mathcal{O}(2^n)$. Despite being exponential, this is the big complexity speedup. We in fact no longer need to look at all the permutations, but we can just consider the subsets. Then, for each iteration we have to invert $k$ matrices $(k-1)\times (k-1)$, which has complexity $\mathcal{O}(k^3)$. Just as before, doing this for each steps results in a complexity of $\mathcal{O}(n^4)$. In conclusion we can manage to go from the brute force complexity $\mathcal{O}(n!n^4)$ to $\mathcal{O}(2^n n^4)$. Regarding space complexity, for the memoization we need two tables, one for the optimal costs, and one for the paths. Each table stores an entry for all possible subsets with keys growing as $\mathcal{O}(n)$, which results in a space complexity of $\mathcal{O}(n2^n)$.
\par
We briefly conclude this section, reminding that for large number of parameters this algorithm \gf{having an exponential trend becomes highly costly}. Therefore one could consider using either heuristics or other techniques such as simulate annealing or genetic algorithms.

% \subsection{Robustness against Sloppiness}
% [about this I'm not sure if we wanted to say something. In this figure I had sampled random QFIM(PSD,symmetric,$n\times n$ matrices), and did a histogram with $\frac{C_{\text{SE}}-C_{\text{JE}}}{C_{\text{JE}}}$. The plots are ugly, but they underline how SE is a procedure which is resistant to sloppiness. And indeed as dimension grows it's more probable to have at least a "bad" parameter(here potentially it may  enter the talk about how with systems with many parameters it's usually formed a hierarchy wich spans many order of magnitute\\
% https://arxiv.org/abs/1303.6738). In any case we may just name this in Sec. \ref{sec:SU22parsqubit} where we notice this.]
% \begin{figure}[h]
%     \centering
%     \includegraphics[width=0.6\linewidth]{Cattura3.PNG}
%     \caption{Caption}
%     \label{fig:enter-label}
% \end{figure}

\section{Two-parameter SU(2) model with pure states}\label{ap:twoparams}
We consider a family of Hamiltonians of the form
\begin{equation}
    H_{B,\theta} = B \big( \cos\theta \, J_x + \sin\theta \, J_z \big) \,,
\end{equation}
where $J_i$ ($i=x,y,z$) are the generators of $SU(2)$ in the spin-$j$ representation, and $\mathbf{n}_\theta = (\cos\theta, 0, \sin\theta)$. 
%We restrict our attention to a particular class of states of the form
%\begin{equation}
 %   \ket{\psi_0} = \cos\alpha \,\ket{j} + e^{i\varphi}\sin\alpha \,\ket{-j},
%\end{equation}
%where $\ket{\pm j}$ are the eigenstates of $J_z$ associated with its extremal eigenvalues.

A general qubit state $\rho_0$ can be written in Bloch vector form as
\begin{equation}
    \rho_0 = \frac{1}{2} \ \mathbb{I} + \mathbf{r}_0 \cdot \boldsymbol{J},
\end{equation}
where $$\textbf{r}_0 = (\text{Tr}[\sigma_x \rho_0],\text{Tr}[\sigma_y \rho_0],\text{Tr}[\sigma_z \rho_0]) $$ with $|\mathbf{r}_0|= 1$ and $\boldsymbol{J} =(J_x, J_y, J_z)=  \frac{1}{2}( \sigma_x, \sigma_y, \sigma_z)$. $\sigma_i$ are the Pauli matrices.  

%A general qutrit state can be written in the generalized Bloch form using normalized Gell-Mann matrices
A general qutrit state can be written through its decomposition on an operator base ${b_i}$
% \begin{equation}
%     \rho_0 = \frac{1}{3} \mathbb{I} + \frac{1}{2} \sum_{i=1}^8 r_i \Gamma_i = \frac{1}{3} \mathbb{I} + \frac{1}{\sqrt{2}} \sum_{i=1}^8 r_i b_i,
% \end{equation}
\begin{equation}
    \rho_0 = \sum\limits_{i=1}^9 r_i b_i 
\end{equation}
where $\mathbf{r} =\{r_i\}= (\text{Tr}[\rho_0 b_1],\dots,\text{Tr}[\rho_0 b_9]) $, which should not be confused with the Bloch vector associated to the qutrit. We note that $r_9 = 1/\sqrt{3}$, and pure states are such that $\text{Tr}[\rho^2]=\left|\textbf{r} \right|^2 = 1$. The elements of the base $b_i$ are just the normalized Gell-Mann matrices $$\textbf{b} =\{b_i\}=\{\Gamma_1/\sqrt{2}, ..., \Gamma_8/\sqrt{2}, \mathbb{I}/\sqrt{3}\}$$ are the Gell-Mann matrices with $\text{Tr}[b_i b_j] = \delta_{ij}$.

%with the first eight components satisfying $|\mathbf{r_{1,\dots,8}}|= 1$ is the qutrit Bloch vector and $$\textbf{b} =\{b_i\}=\{\Gamma_1/\sqrt{2}, ..., \Gamma_8/\sqrt{2}, \mathbb{I}/\sqrt{3}\}$$ are the Gell-Mann matrices with $\text{Tr}[b_i b_j] = \delta_{ij}$.

For a probe state $\rho_0$ evolving under the above Hamiltonian for a time $t$, the elements of the QFIM are given by
\begin{align}
    Q_{BB} &= 4 t^2 \big[ \braket{J_{\mathbf{n}_\theta}^2}_0 - \braket{J_{\mathbf{n}_\theta}}_0^2 \big],\\
    Q_{\theta\theta} &= 16 \sin^2\!\frac{Bt}{2} \big[ \braket{J_{\mathbf{n}_1}^2}_0 - \braket{J_{\mathbf{n}_1}}_0^2 \big],\\
    %\label{eq:3Qsu2qutrit}
    Q_{B\theta} &= -4 t \sin\!\frac{Bt}{2} \big[ \braket{\{J_{\mathbf{n}_1}, J_{\mathbf{n}_\theta}\}}_0 - 2 \braket{J_{\mathbf{n}_1}}_0 \braket{J_{\mathbf{n}_\theta}}_0 \big],
\end{align}
where $$\textbf{n}_1 = \left(\cos\frac{Bt}{2}\sin\theta,-\sin\frac{Bt}{2},-\cos\frac{Bt}{2}\cos\theta \right)$$ is the derivative direction with respect to $\theta$. For the Uhlmann matrix element relevant to compatibility conditions, we have
\begin{equation}
    D_{\theta B} = 4 t \sin\!\frac{Bt}{2} \, \braket{J_{\mathbf{n}_2}}_0,
\end{equation}
where $$\textbf{n}_2 = \left(\sin\frac{Bt}{2}\sin\theta,\cos\frac{Bt}{2},-\sin\frac{Bt}{2}\cos\theta\right).$$

These formulas apply to both qubit ($j=1/2$) and qutrit ($j=1$) systems; the difference lies in the explicit form of the spin operators $J_i$. For the qubit case, they are expressed in terms of Pauli matrices as
\[
J_x = \frac{1}{2}\sigma_x,\quad
J_y = \frac{1}{2}\sigma_y,\quad
J_z = \frac{1}{2}\sigma_z,
\]
where $\sigma_i$ are Pauli matrices. For the qutrit case, the generators take the form
$$
J_x = \frac{1}{\sqrt{2}}
\begin{pmatrix}
    0 & 1 & 0\\
    1 & 0 & 1\\
    0 & 1 & 0
\end{pmatrix},\quad
J_y = \frac{1}{\sqrt{2}}
\begin{pmatrix}
    0 & -i & 0\\
    i & 0 & -i\\
    0 & i & 0
\end{pmatrix},\quad
J_z =
\begin{pmatrix}
    1 & 0 & 0\\
    0 & 0 & 0\\
    0 & 0 & -1
\end{pmatrix}.
$$

\bibliographystyle{elsarticle-num}
\bibliography{bib}

\end{document}